\documentclass{svmult}

\usepackage{mathptmx}       % selects Times Roman as basic font
\usepackage{helvet}         % selects Helvetica as sans-serif font
\usepackage{courier}        % selects Courier as typewriter font
\usepackage{type1cm}        % activate if the above 3 fonts are
\usepackage{ulem}   % package for \sout{...} and \xout{...} to delete text
\usepackage{color}  % package for \textcolor{"color"}{...}
\usepackage{amsmath}
\usepackage{bm}
\usepackage[numbers,sort&compress]{natbib}
\usepackage{makeidx}         % allows index generation
\usepackage{graphicx}        % standard LaTeX graphics tool
\usepackage{multicol}        % used for the two-column index
\usepackage[bottom]{footmisc}% places footnotes at page bottom

\makeindex             % used for the subject index
\begin{document}

\title*{Average-Atom Model for X-ray Scattering from Warm Dense Matter}
\titlerunning{X-ray Scattering from WDM}% for an abbreviated version of
\author{W. R. Johnson, J. Nilsen and K. T. Cheng}
\institute{W. R. Johnson \at University of Notre Dame, Notre Dame, IN 46556,
\email{johnson@nd.edu}
\and J. Nilsen \at Lawrence Livermore National Laboratory, Livermore, CA 94551
\email{nilsen1@llnl.gov}
\and K. T. Cheng \at Lawrence Livermore National Laboratory, Livermore, CA 94551
\email{ktcheng@llnl.gov}
}
\maketitle

\abstract{A scheme for analyzing Thomson scattering of x-rays by warm dense matter,
based on the average-atom model, is developed.
Emphasis is given to x-ray scattering by bound electrons.
Contributions to the scattered x-ray spectrum from elastic scattering
by electrons moving with
the ions and from inelastic scattering by free and bound electrons
are evaluated using parameters
(chemical potential, average ionic charge, free electron density, bound and
continuum wave functions, and occupation numbers) taken from the average-atom model.
The resulting scheme provides a relatively simple diagnostic for use in connection
with x-ray scattering measurements.
Applications are given to dense hydrogen, beryllium, aluminum,
titanium, and tin plasmas. At high momentum transfer,
contributions from inelastic scattering by bound electrons are dominant features
of the scattered x-ray spectrum for aluminum, titanium, and tin.}

\section{Introduction}

Measurements of Thomson scattering of x-rays provide information on
temperatures, densities and ionization balance in warm dense matter.
Various techniques for inferring  plasma properties from x-ray scattering
measurements have been developed over the past decade
\cite{GGL:02,LB:02,GG:03,LM:03,GGL:03,GGL:03a,HR:04,GGL:06,GL:07,TB:08,KN:08,
FT:09,DL:09,KG:09,GL:10,MM:10,TB:10,RR:10,KD:11,VD:12,FL:12,ZB:12}. Many of
these techniques, together with the underlying theory, were reviewed
by Glenzer and Redmer in Ref.~\cite{GR:09}.

The present average-atom scheme is based on a theoretical description
of x-ray scattering proposed by Gregori \cite{GG:03}, the important difference
being that parameters used to evaluate the Thomson-scattering dynamic
structure function are taken from the average-atom model.
The average-atom model used here was introduced in Ref.~\cite{JGB:06} to study
electromagnetic properties of plasmas.
The scheme developed here to analyze Thomson scattering is closely related
to that of Sahoo et al.~\cite{SG:08},
where a somewhat different version of the average-atom model was used.
Predictions from the present model differ substantially from those
in Ref.~\cite{SG:08}. The origin and consequences of these differences
will be  discussed later.

The Thomson scattering cross section for an incident photon with energy,
momentum ($\hbar \omega_0, \, \hbar {\bm k}_0$),
and polarization ${\bm \epsilon}_0$ scattering to a state with energy,
momentum ($\hbar \omega_1, \, \hbar {\bm k}_1$),
and polarization ${\bm \epsilon}_1$ is
\begin{equation}
 \frac{d\sigma}{d\hbar\omega_1 d\Omega} = \left(  \frac{d\sigma}{d\Omega}
   \right)_{\!\text{Th}}\frac{\omega_1}{\omega_0}\
   S(k,\omega), \label{eq1}
\end{equation}
where
\begin{equation}
 \left(  \frac{d\sigma}{d\Omega} \right)_{\!\text{Th}} = \,
 |{\bm \epsilon}_0\cdot{\bm \epsilon}_1|^2 \left( \frac{e^2}{mc^2}
 \right)^{\!\!2}.
\end{equation}
The {\it dynamic structure function} $S(k,\omega)$ appearing in Eq.~(\ref{eq1})
depends on two variables: $k = |{\bm k}_0 - {\bm k}_1|$ and
$\omega=\omega_0 - \omega_1$.
As shown in the seminal work of Chihara \cite{CH:87,CH:00}, $S(k,\omega)$ can be
decomposed into three terms: the first $S_{ii}(k,\omega)$ is the contribution
from elastic scattering by electrons that follow the ion motion,
the second $S_{ee}(k,\omega)$ is the contribution from scattering by
free electrons, and the third  $S_b(k,\omega)$ is the contribution from
bound-free transitions (inelastic scattering by bound electrons) modulated
by the ionic motion. The modulation factor is ignored here
when evaluating the bound-free contribution.
For the bound-free scattering, calculations are carried out using both
average-atom final states and plane-wave final states.
Substantial differences are found between average-atom
and plane-wave calculations, particularly in the low-momentum transfer region
of the scattered x-ray spectrum.

The average-atom model is discussed in Sec.~\ref{avat} followed by
a discussion of the three contributions to the structure functions in
Sec.~\ref{dynam}. In Sec.~\ref{apps}, applications are given to hydrogen,
beryllium, aluminum, titanium, and tin plasmas.

\section{Average-Atom Model\label{avat}}

The average-atom model is a quantum mechanical version of the
temperature-dependent Thomas-Fermi model of a plasma developed
sixty-three years ago by Feynman, Metropolis and Teller~\cite{FMT:49}.
In this model, the plasma is divided into neutral Wigner-Seitz (WS) cells
(volume per atom $V_{\scriptscriptstyle WS} = A/\rho N_{\rm A}$,
where $A$ is the atomic weight, $\rho$ is the mass density, and
 $N_{\rm A}$ is Avogadro's number).
Inside each WS  cell is a nucleus of charge $Z$ and $Z$ electrons.
Some of these electrons are in bound states and some in continuum states.
The continuum density is finite at the cell boundary and merges into the
uniform free-electron density $n_e=Z_f/V_{\scriptscriptstyle WS}$
outside the cell, where $Z_f$ is the number of free electrons per ion.
Each neutral cell can, therefore, be regarded
as an ion imbedded in a uniform sea of free electrons.
To maintain overall neutrality, it is necessary to introduce a uniform
(but \mbox{inert}) positive background density $Z_f/V_\text{\tiny WS}$.
The model, therefore, describes an isolated (neutral) ion floating in a
(neutral) ``jellium'' sea.

%----- figure 1 -----
\begin{figure}[t]
%\vspace{150pt}
\centering
\includegraphics[scale=0.6]{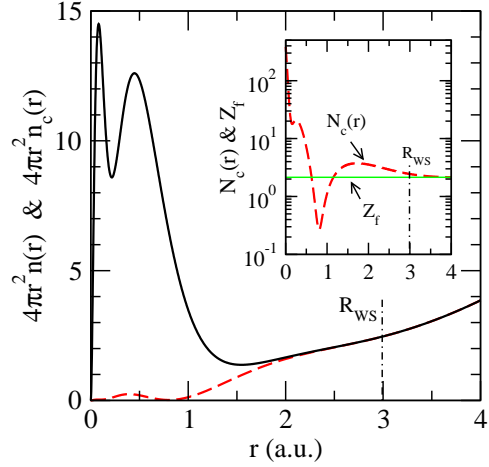}
\caption{Main plot: The radial density $4\pi r^2 n(r)$ in a.u.\
for metallic Al at $k_{\scriptscriptstyle B} T=5$~eV (solid black curve)
integrates to $Z=13$ for $r \leq R_\text{\tiny WS}=2.991~a_0$.
The continuum contribution $4\pi r^2 n_c(r)$ (dashed red curve),
also in a.u.,\ integrates to 3 for $r \leq R_\text{\tiny WS}$.
The bound $1s$, $2s$, and $2p$ shells
are completely occupied at this temperature.
Inset: The dashed red curve illustrates the Friedel oscillations of
the continuum density and shows how $N_c(r) = n_c(r) V_{\text{\tiny WS}}$
converges to $Z_f = n_e V_{\text{\tiny WS}}$ (solid green line) for
$r>R_\text{\tiny WS}$. The chemical potential predicted by the model is
$\mu = 0.2406$~a.u.\ and the number of free electrons per ion is $Z_f = 2.146$.
\label{fig1}}
\end{figure}
%-----------

The quantum-mechanical model here, which is discussed in Ref.~\cite{JGB:06},
is a nonrelativistic version of the {\it Inferno} model of
Liberman \cite{DL:79} and the more recent {\it Purgatorio} model of
Wilson et al.~\cite{BS:06};
it is similar to the nonrelativistic average-atom model described
by Blinski and Ishikawa~\cite{BI:95}. Specifically, each electron
in the ion is assumed
to satisfy the central-field Schr\"{o}dinger equation
\begin{equation}
 \left[ \frac{p^2}{2} -\frac{Z}{r} + V \right] \psi_a({\bm r})
= \epsilon_a\, \psi_a({\bm r}) , \label{sq1}
\end{equation}
where $a=(n,l)$ for bound states or $(\epsilon,l)$ for continuum states.
Atomic units (a.u.)\ where $e=\hbar=m=4\pi\epsilon_0=1$ are used here.
In particular, 1 a.u.\ in energy equals 2 Rydbergs (27.211 eV),
and 1 a.u.\ in length equals 1 Bohr radius $a_0$ (0.529 \AA).

The wave function $\psi_a({\bm r})$ is decomposed in a spherical basis as
\begin{equation}
  \psi_a({\bm r}) = \frac{1}{r} P_a(r)\, Y_{l_am_a}(\hat{r})\, \chi_{\sigma_a},
\end{equation}
where $Y_{lm}(\hat{r})$ is a spherical harmonic and
$\chi_{\sigma}$ is a two-component electron spinor.
The bound and continuum radial functions $P_a(r)$ are normalized as
\begin{align}
\int_0^\infty\!\!  dr\, P_{nl}(r)\, P_{n'l}(r) &={} \delta_{nn'}, \\
\int_0^\infty\!\!  dr\, P_{\epsilon l}(r)\, P_{\epsilon' l}(r) &={}
\delta(\epsilon-\epsilon') ,
\end{align}
respectively. The central potential $V(r)$ in Eq.~(\ref{sq1})
is taken to be the self-consistent Kohn-Sham potential \cite{kohnsham}
\begin{equation}
V(r) = 4\pi \!\int\! \frac{1}{r_>} \, r'^2 \, n(r') \, dr'
-  \bigg[ \frac{3}{\pi} \, n(r) \bigg]^{\!\frac{1}{3}},
\label{sq5}
\end{equation}
where the first term in the right-hand side % of Eq.\ (\ref{sq5})
is the direct screening potential with $r_> = \text{max}(r,\,r')$
and the second term is the Kohn-Sham exchange potential.
While electron-electron
interactions inside the Wigner-Seitz cells are reasonably well
accounted for by this simple model, it should be noted that
eigenvalues in the Kohn-Sham potential are poor approximations to
ionization energies, leading to inaccurate thresholds and peaks of
bound-free contributions
to $S(k,\omega)$, which can differ from experiment by 20 -- 30\%.
The electron density $n(r)$ in Eq.~(\ref{sq5}) has contributions from
bound states $n_b(r)$ and from continuum states $n_c(r)$,
\begin{equation}
 n(r) = n_b(r)+n_c(r).
 \end{equation}
The bound-state contribution to the density $n_b(r)$ is
\begin{equation}
  4\pi r^2 n_b(r) = \sum_{nl} \frac{2(2l+1)}
  {1+ \exp[(\epsilon_{nl}-\mu)/k_{\scriptscriptstyle B} T]}\,
   P_{nl}(r)^2 , \label{sq3}
\end{equation}
where $\epsilon_{nl}$ is the bound-state energy, $\mu$ is the chemical
potential, and the sum over $(n,l)$ ranges over all bound subshells.
The continuum contribution to the density $n_c(r)$ is given by
\begin{equation}
  4\pi r^2 n_c(r) = \sum_{l}\int_0^\infty \!\! d\epsilon\, \frac{ 2(2l+1) }
  {1+ \exp[(\epsilon-\mu)/k_{\scriptscriptstyle B} T]}\,
   P_{\epsilon l}(r)^2 . \label{sqex}
\end{equation}
Finally, the chemical potential $\mu$ is chosen to
ensure
charge neutrality inside the WS cell:
\begin{equation}
Z = \int_{r \le R_\text{\tiny WS}}\!\! n(r)\, d^3r\equiv
\int_0^{R_\text{\tiny WS}} \!\! 4\pi r^2 n(r)\, dr \ . \label{sq6}
\end{equation}
Equations (\ref{sq1}--\ref{sq6}) above are solved self-consistently
to give the chemical potential $\mu$, the potential energy function $V(r)$
and the electron density $n(r)$.

The upper limit in Eq.~(\ref{sq3}) is determined by systematic trial and error.
The values of $n$ and $l$ are increased starting from $n=1$ and $l=0$.
If a state is bound, it is included in the sum, otherwise not.
At metallic densities and temperatures below 100~eV (warm dense matter)
fewer than a dozen states typically bind.
To carry out the sum-integral in Eq.~(\ref{sqex}) for the
continuum density, we typically use 12 partial waves ($l$) and 40 to 50 energy
points ($\epsilon$) for each partial wave.
The energy grid for the integral in Eq.~(\ref{sqex}) is chosen using
a modified Gauss-Laguerre scheme.
Thus, one is faced with solving a system of roughly 500 coupled second-order
differential equations. These equations are solved iteratively using a
predict-correct scheme based on Adam's method \cite[Chap.~2.3]{book}.

The boundary conditions used in solving Eq.~(\ref{sq1}) deserve some mention.
Bound-state wave functions and their derivatives are matched at the boundary
$r=R_\text{\tiny WS}$ to solutions outside the WS sphere (where $V=0$)
that vanish exponentially as $r\to \infty$.
Similarly, continuum functions and their derivatives are matched to
phase-shifted free-particle wave functions at $r=R_\text{\tiny WS}$.
It should be noticed that the continuum density $n_c(r)$ inside the WS sphere,
which oscillates as predicted by Freidel~\cite{JF:54},
is distinctly different from the uniform free electron density $n_e$.
In the present model, $n_c(r)$ smoothly approaches $n_e$ outside the sphere.
These points are illustrated in Fig.~\ref{fig1}, where the bound-state and
continuum densities are plotted for Al at metallic
density and temperature $k_{\scriptscriptstyle B} T = 5$~eV.
Occupation numbers of bound states
and continuum partial-wave states inside
the WS sphere are given along with bound-state eigenvalues in Table~\ref{tab1}.

%----- table 1 -----
\begin{table}[t]
\centering
\caption{Aluminum at density 2.70~g/cc and $k_{\scriptscriptstyle B} T=5$~eV.
Bound-state and continuum partial-wave occupation numbers
inside the WS sphere are given, along with bound-state eigenenergies.
The Ne-like core is seen to be almost completely occupied.
The sum of the bound-state and continuum occupation numbers
is precisely $Z=13$.
 \label{tab1}}
\begin{tabular}{c @{\hspace{1pc}}c @{\hspace{1pc}}r @{}c@{\hspace{2pc}}
                c @{\hspace{1pc}}c}
\hline\hline
  \multicolumn{3}{c}{Bound States} &  % \rule[2ex]{2em}{0pt}
& \multicolumn{2}{c}{Continuum} \\
\cline{1-3} \cline{5-6}
State & occ\# & $\epsilon$(eV)~~ && $l$ & occ\# \\
\hline
   $1s$ & 2.0000   &   -1485.07  &&      0  &     0.9130\\
   $2s$ & 2.0000   &     -92.16  &&      1  &     1.3263\\
   $2p$ & 5.9998   &     -54.87  &&      2  &     0.6192\\
        &          &             &&      3  &     0.1173\\
        &          &             &&      4  &     0.0209\\
        &          &             &&      5  &     0.0031\\
        &          &             &&      6  &     0.0004\\
        &          &             &&      7  &     0.0000\\
\hline
  $N_b$ & 9.9998   &             &&   $N_c$ &     3.0002\\
\hline\hline
\end{tabular}
\end{table}
%----------

The boundary conditions used here differ from those used
by Sahoo et al.\ in Ref.~\cite{SG:08},
where the first derivative of the wave function is required to vanish at
$R_\text{\tiny WS}$. The differences in boundary conditions lead to major
differences in the average-atom structure. For example, the model used
in Ref.~\cite{SG:08} predicts that the $M$ shell of metallic Al is partially
occupied at temperatures $k_{\scriptscriptstyle B} T\leq 10$~eV,
whereas the present model predicts that the $M$ shell is empty in this
temperature range. Consequences of such differences are discussed later
in Sec.~\ref{apps}.

\section{Dynamic Structure Function\label{dynam}}

As noted in the introduction, the theoretical model developed by
Gregori et al.~\cite{GG:03}, with input from the average-atom model,
is used here to evaluate the dynamic structure function $S(k,\omega)$.
The ion-ion contribution $S_{ii}(k,\omega)$ is evaluated in terms of
Fourier-transforms of electron densities and % formulas for
the static ion-ion structure function $S_{ii}(k)$.
Formulas for $S_{ii}(k)$ are given in Ref.~\cite{AD:98}.
Here, we use modified formulas that include options
discussed in Ref.~\cite{GGL:06} for different electron and ion temperatures.
The dominant effect of different electron and ion temperatures is to modify
the relative size of elastic to inelastic contributions to $S(k,\omega)$.
The use of different temperatures is, therefore, a convenient tool for
fitting experimental data, even in cases where equilibrium is
reached.
The inability to fit experimental data in some equilibrium cases without
such an artifice is a weakness in the present scheme and is the subject
of current research. The electron-electron contribution
$S_{ee}(k,\omega)$ is expressed in terms of the dielectric function
$\epsilon(k,\omega)$ of the free electrons which in turn is evaluated
using the random-phase approximation (RPA). Plasmon resonances are present
in $S_{ee}(k,\omega)$ at low momentum transfers $k$.
Finally, bound-state contributions to the dynamic structure function are
evaluated using average-atom bound-state wave functions for the initial state.
The final-state continuum wave function is described in two different
ways: (1) approximating the final-state by a plane wave as in Ref.~\cite{SG:08},
and (2) using an average-atom final-state that approaches a
plane wave asymptotically. There are dramatic differences between these two
choices, especially at low momentum transfers. The more realistic average-atom
choice automatically includes ionic Coulomb-field effects.

\subsection{Ion-Ion Structure Function}

The contribution to the dynamic structure function from elastic scattering by
electrons following the ion motion $S_{ii}(k,\omega)$ is expressed in terms of
the corresponding static ion-ion structure function $S_{ii}(k)$ as:
\begin{equation}
S_{ii}(k,\omega) = |f(k)+ q(k)|^2 \, S_{ii}(k)\, \delta(\omega) .
\label{eqsii}
\end{equation}
In the above, $f(k)$ is the Fourier transform of the bound-state density
and $q(k)$ is the Fourier transform of the density of electrons that screen
the ionic charge.
In the average-atom approximation, the screening electrons are the continuum
electrons inside the Wigner-Seitz sphere and
\begin{equation}
f(k)+q(k) = 4 \pi \!\int_0^{R_{\scriptscriptstyle WS}} \!\! r^2
\left[ n_b(r)+n_c(r) \right] j_0(kr) \, dr ,
\end{equation}
where $j_0(z)$ is a spherical Bessel functions of order $0$.
Note that $f(0)+q(0) = Z$ in the average-atom model.
In the applications discussed later, $\delta(\omega)$  is replaced by
an ``instrumental'' Gaussian, with full-width at half
maximum = 10 eV. This value is chosen because typical experiments in Be \cite{DL:09}
utilize a spectrometer with a 10-eV instrument width and use a Cl Ly-$\alpha$ source
at 2.96 keV.

Approximate schemes to evaluate the static structure functions $S_{ii}(k)$
are discussed, for example, in Ref.~\cite{JPH:06}.
Here, we follow Gregori et al.~\cite{GG:03} and make use of formulas given by
Arkhipov and Davletov~\cite{AD:98}
that account for both quantum-mechanical and screening effects.
The function $S_{ii}(k)$ in Ref.~\cite{AD:98} is expressed
in terms of the
Fourier transform of the ion-ion interaction potential $\Phi_{ii}(r)$
through the relation:
\begin{equation}
S_{ii}(k) = 1 - \frac{n_i}{k_{\scriptscriptstyle B} T} \Phi_{ii}(k).
\end{equation}

\subsubsection*{Different Electron and Ion Temperatures}

\begin{figure}[t]
\centering
\includegraphics[scale=0.52]{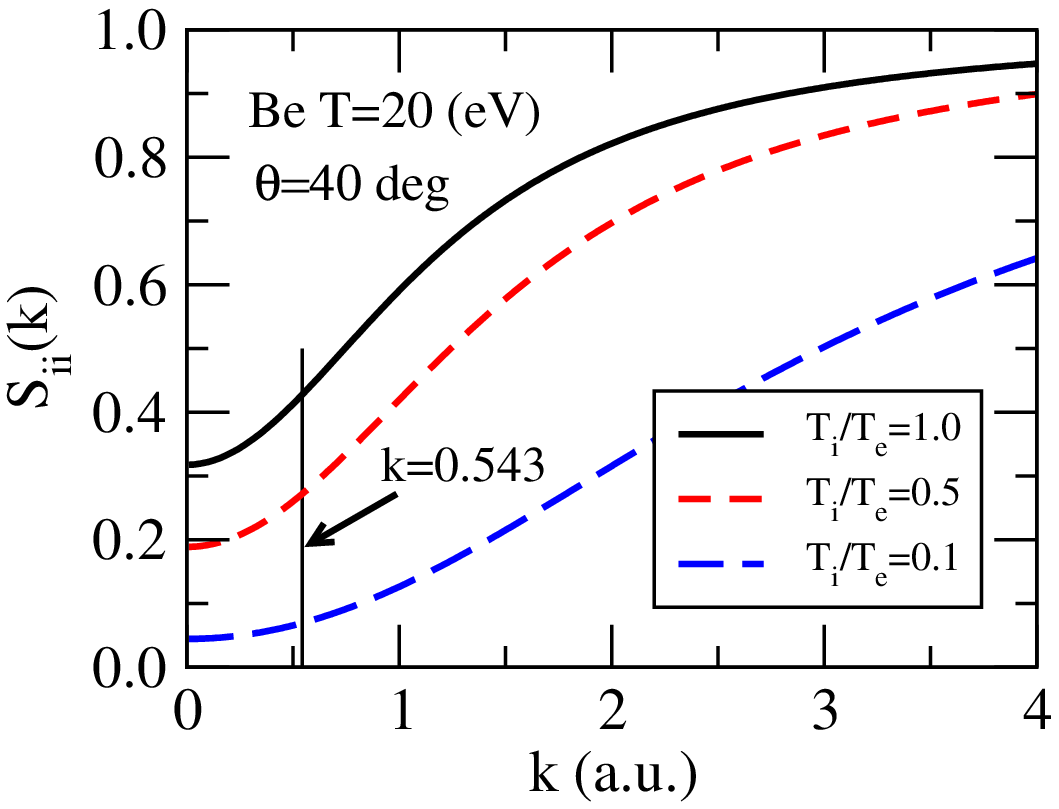}~
\includegraphics[scale=0.52]{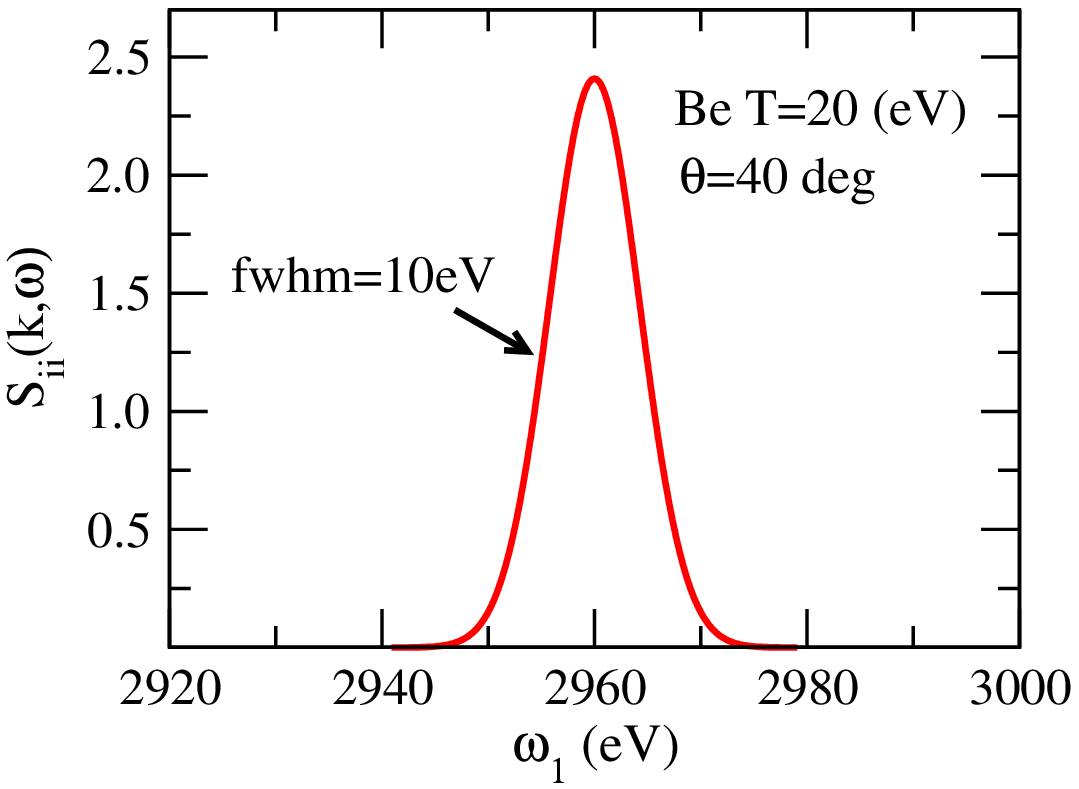}
\caption{Left panel: $S_{ii}(k)$ is shown for Be metal at electron temperature
$k_{\scriptscriptstyle B}T_e=20$ eV and ion-electron temperature ratios
$T_i/T_e=(1,\ 0.5,\ 0.1)$ illustrated in solid, short-dashed and
long dashed curves , respectively.
The value $k$ = 0.543 corresponds to an incident photon $\omega_0=2960$~eV
scattered at angle 40$^\circ\!$.
Right panel: $S_{ii}(k,\omega)$ in a.u.\ for Be metal
at $k_{\scriptscriptstyle B}T_e=20$ eV and $T_i/T_e =0.1$,
where the function $\delta(\omega)$ is
replaced by a Gaussian of width 10~eV and $\omega_1 = \omega_0 - \omega$
is the energy of the scattered x-ray.
\label{fig2}}
\end{figure}

In the average-atom model, $T$ is the electron temperature
$T_e$ which, in equilibrium, is equal to the ion temperature $T_i$.
To allow for different electron and ion temperatures, the equations for
$S_{ii}(k)$ given by Arkhipov and Davletov~\cite{AD:98} are modified following
the prescription laid out by Gregori et al.~\cite{GGL:06}.
The electron temperature $T_e$ is replaced by
an effective temperature $T_e^\prime$ that accounts for degeneracy effects
at temperatures lower than the Fermi temperature $T_F$. Similarly, the
ion temperature $T_i$ is replaced by an effective temperature $T_i^\prime$
that accounts for ion degeneracy effects at temperatures lower than the ion
screened Debye temperature $T_D$. Explicit formulas for $S_{ii}(k)$ are
found in Ref.~\cite{GGL:06}.
The dramatic effect of different electron and ion
temperatures on the static structure functions $S_{ii}(k)$ for metallic Be
at $T_e=20$~eV are illustrated in the left panel of Fig.~\ref{fig2}.
This figure is similar to the upper-left panel of Fig.~1 in Ref.~\cite{GGL:06},
which was obtained under similar condition.
In the right panel of Fig.~\ref{fig2}, contributions to $S_{ii}(k,\omega)$
for Be at $T_e=20$~eV and  $T_i=2$~eV are shown.

%----- figure 2 -----
%-----------

\subsection{Electron-Electron Structure Function}

The electron-electron structure function $S_{ee}(k,\omega)$ is expressed in terms
of the free-electron dielectric function $\epsilon(k,\omega)$ through the relation Eq.~(42) in
Ref.~\cite{CH:00}:
\begin{equation}
S_{ee}(k,\omega) = - \frac{1}{1-\exp(-\omega/k_{\scriptscriptstyle B} T)} \frac{k^2}{4\pi^2 n_e}
\text{Im}\left[\frac{1}{\epsilon(k,\omega)}\right]. \label{see}
\end{equation}

We follow Gregori et al.~\cite{GG:03} and evaluate the dielectric function in the
random-phase approximation (RPA):
\begin{multline}
\epsilon(k,\omega) = 1 + \frac{4}{\pi k^2} \int_0^\infty \!\!{\mathcal{F}}(p)\, p^2 dp
\\ \times
\int_{-1}^1 d\mu
\left[\frac{1}{k^2-2p k \mu + 2\omega + i\nu} + \frac{1}{k^2+2p k \mu - 2\omega - i\nu} \right] ,
\label{eq16}
\end{multline}
where
\begin{equation}
{\mathcal{F}}(p) = \frac{1}{1+\exp[(p^2/2-\mu)/k_{\scriptscriptstyle B} T]}
\end{equation}
is the free-electron Fermi distribution. The parameter $\nu$ in Eq.~{\ref{eq16}}
is a positive infinitesimal.
The chemical potential $\mu$ is obtained from
the average-atom model.
It should be noted that the RPA dielectric function reduces to
the Lindhard dielectric function \cite{Li:54,Ma:00} in the limit
$k_{\scriptscriptstyle B} T\to 0$, where ${\mathcal{F}}(p)$ reduces to
a unit step function that vanishes for $p>k_{\scriptscriptstyle F}$.
In the finite $T$ case, we find
\begin{align}
\int_{-1}^1
\frac{d\mu}{k^2-2p k \mu + 2\omega  + i\nu} &=\  \frac{1}{2pk}\,
\ln\!\left[\frac{k^2+2pk+2\omega+i\nu}{k^2-2pk+2\omega+i\nu}\right] \\
\int_{-1}^1
\frac{d\mu}{k^2+2p k \mu - 2\omega  - i\nu} &=\ \frac{1}{2pk}\,
\ln\!\left[\frac{k^2+2pk-2\omega-i\nu}{k^2-2pk-2\omega-i\nu}\right] .
\end{align}
From the above expressions, it is clear that $\text{Re}[ \epsilon(k,\omega)]$
is an even function of $\omega$ and that
$\text{Im}[\epsilon(k,\omega)]$ is an odd function of $\omega$. Therefore,
we can limit ourselves to cases where
$\omega$ is positive.  For $\omega>0$, one obtains in the limit $\nu \to 0+$,
\begin{multline}
\lim_{\nu \to 0} \left\{ \ln\left[\frac{k^2+2pk+2\omega+i\nu}{k^2-2pk+2\omega+i\nu}\right] +
 \ln\left[\frac{k^2+2pk-2\omega-i\nu}{k^2-2pk-2\omega-i\nu}\right] \right\} \\
 =  \ln \left|\frac{k^2+2pk+2\omega}{k^2-2pk+2\omega}\right| +
 \ln\left|\frac{k^2+2pk-2\omega}{k^2-2pk-2\omega}\right| \\
 + i\pi\ \text{only in the interval}\ |k^2-2\omega|\leq 2pk \leq k^2+2\omega .
\end{multline}
 It follows that
\begin{multline}
\text{Re}[\epsilon(k,\omega)] = 1 + \frac{2}{\pi k^3} \int_0^\infty {\mathcal{F}}(p)\, p\, dp
\\ \times
 \left[ \ln \left|\frac{k^2+2pk+2\omega}{k^2-2pk+2\omega}\right| +
 \ln\left|\frac{k^2+2pk-2\omega}{k^2-2pk-2\omega}\right| \right]
 \end{multline}
 and
 \begin{equation}
 \text{Im}[\epsilon(k,\omega)]  =  \frac{2}{k^3} \int_a^b {\mathcal{F}}(p)\,p\,dp
                          = \frac{2 k_{\scriptscriptstyle B} T}{k^3} \log\left[
                         \frac{1+\exp[(\mu-a^2/2)/k_{\scriptscriptstyle B} T]}
                         {1+\exp[(\mu-b^2/2)/k_{\scriptscriptstyle B} T]}\right]
\end{equation}
with $a = |2\omega-k^2|/2k$ and $b=(2\omega+k^2)/2k$.
From Eq.~(\ref{see}) and the fact that $\text{Im}[1/\epsilon(k,\omega)]$
is an odd function of $\omega$, it follows that
\begin{equation}
S_{ee}(k,\omega) = \exp(\omega/{k_{\scriptscriptstyle B}T}) S_{ee}(k,-\omega).
\end{equation}
Therefore, in the absence of bound-state contributions, measuring the cross-section ratio
at down-shifted and up-shifted plasmon peaks, provides an experimental method for
determining the plasma temperature.

%----- figure 3 -----
\begin{figure}[t]
\centering
\includegraphics[scale=0.8]{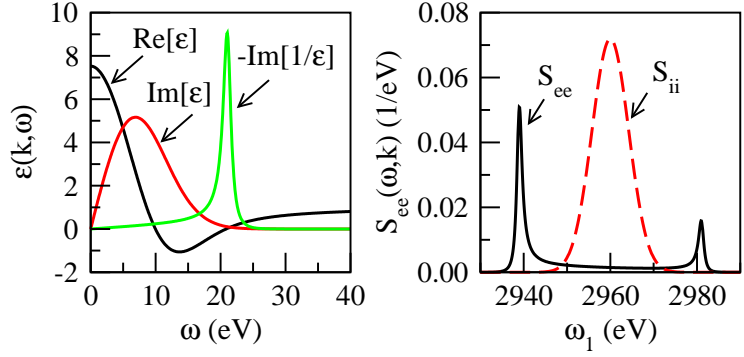}
\caption{Left panel: Real and Imaginary parts of the dielectric function
$\epsilon(k,\omega)$ are plotted along
with $-\text{Im}[1/\epsilon]$
for coherent scattering of a 2960-eV photon at 20$^\circ\!$ from Be metal at
$k_{\scriptscriptstyle B} T = 18$~eV.
 Right panel: The resulting electron-electron structure function
$S_{ee}(k,\omega)$ (solid line) is shown
together with $S_{ii}(k,\omega)$ (dashed line) which has an
instrumental width of 10~eV. In this example $k=0.2757$~a.u.\ and
$\omega_1 = 2960 - \omega$ (eV) is the energy of the scattered x-ray.
\label{fig3}}
\end{figure}
%-----------

The real and imaginary parts of $\epsilon(k,\omega)$,
along with $-\text{Im}[1/\epsilon(k,\omega)]$, are shown
in the left panel of Fig.~\ref{fig3} for Be metal at temperature
$k_{\scriptscriptstyle B} T = 18$~eV. A plasmon peak, associated with
coherent scattering of the incident x-ray by the electrons in the plasma,
is seen near $\omega=20$~eV, where the real part of the dielectric function vanishes.
The corresponding structure function $S_{ee}(k,\omega)$, illustrating
up- and down-shifted plasmon features, is shown in the right-hand panel.
For comparison purposes, the elastic-scattering
structure function $S_{ii}(k,\omega)$ is also shown in the right-hand panel.
The functions displayed in this plot
correspond to scattering of an incident 2960-eV photon at an angle 20$^\circ\!$.
The classical plasma frequency for this case
($n_e = 2.05\times 10^{23}$ cc$^{-1}$)
is $\omega_\text{pl} = 16.8$ eV.

The amplitude and width of plasmon peaks is governed by the coherence parameter
$\alpha = 1/(\lambda_s k)$, defined in Eqs.~(5-7) of Ref.~\cite{GR:09}.
Here, $\lambda_s$ is the shielding length given by
\begin{equation}
\lambda_s = \sqrt{
\frac{k_{\scriptscriptstyle B}T F_{1/2}(\mu/{k_{\scriptscriptstyle B}T})}
{4 \pi n_e F_{-1/2}(\mu/{k_{\scriptscriptstyle B}T})}} \, ,
\end{equation}
where $F_j(x)$ is a complete Fermi-Dirac integral,
\begin{equation}
F_\nu(x) = \frac{1}{\Gamma(1+\nu)} \int_0^\infty\! \! dy\, \frac{y^\nu}{1+\exp(y-x)} \, .
\end{equation}

\begin{figure}[t]
\centering
\includegraphics[scale=0.8]{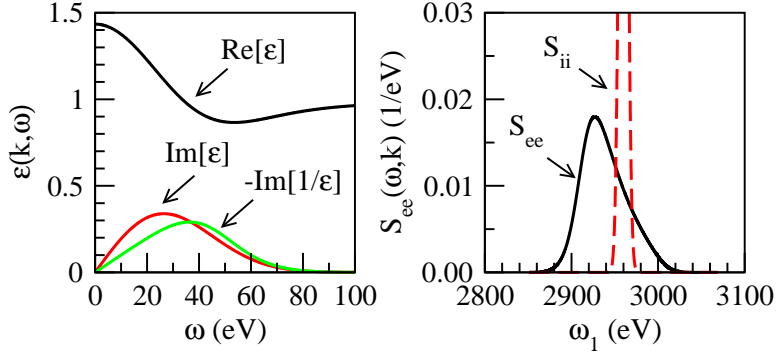}
\caption{Left panel: Real and Imaginary parts of the dielectric function
$\epsilon(k,\omega)$ are plotted along
with $-\text{Im}[1/\epsilon]$
for incoherent scattering of a 2960-eV photon at 90$^\circ\!$ from Be metal
at $k_{\scriptscriptstyle B} T = 18$~eV.
 Right panel: The electron-electron structure function
$S_{ee}(k,\omega)$ (solid line) is shown
together with $S_{ii}(k,\omega)$ (dashed line) which is assumed to have
an instrumental width of 10~eV.
The corresponding momentum transfer is $k=1.123$ a.u.\ and
$\omega_1 = 2960 - \omega$ (eV) is the energy of the scattered x-ray.
\label{fig4}}
\end{figure}

For the example shown in Fig.~\ref{fig3}, the screening length $\lambda_s=1.440$
and the corresponding coherence parameter $\alpha=2.520$, so one expects and
observes plasmon resonances.
At values of $\alpha$ less than than one, coherent
scattering by the plasma no longer occurs and the plasmon peak
in the dielectric function disappears.
An example of this behavior is illustrated in Fig.~\ref{fig4}, where the
dielectric function for scattering of a 2960-eV photon at 90$^\circ\!$
from Be metal at $k_{\scriptscriptstyle B} T=18$~eV is illustrated.
The screening radius in this case remains unchanged $\lambda_s=1.440$,
but the momentum transfer increases to $k=1.123$
and the resulting coherence parameter is $\alpha =0.6188$. Therefore,
as expected, all signs of plasmon resonances are absent in $S_{ee}(k,\omega)$.

%----- figure 4 -----
%-----------

\subsection{Bound-Free Structure Function}

The contribution to the dynamic structure function for Thomson scattering
from ionic bound states $S_b(k,\omega)$
is the sum over contributions from individual subshells
with quantum numbers $(n,l)$:
\begin{align}
S_b(k,\omega)    &=\ \sum_{nl} S_{nl}(k,\omega) \\
S_{nl}(k,\omega) &=\ \frac{o_{nl}}{2l+1} \sum_{m} \int \frac{p\, d\Omega_p}{(2\pi)^3}\,
\left| \int\!\! d^3 r\, \psi^\dagger_{\bm p}(\bm{r})\, e^{i\bm{k}\cdot\bm{r}}\,
\psi_{nlm}(\bm{r})
\right|^2_{\epsilon_p=\omega+\epsilon_{nl}}
 ,\label{snl}
\end{align}
where $o_{nm}$ is the fractional occupation number of subshell $(n,l)$,
and ${\bm k}={\bm k}_0 - {\bm k}_1$ and $\omega = \omega_0-\omega_1$
are the momentum and energy transfers, respectively,
from the incident to the scattered photon as defined in Eq.~(\ref{eq1}).
In the above, $\psi_{\bm p}(\bm{r})$ is an average-atom wave function that
approaches a plane wave $e^{i\bm {p\cdot r}}$ asymptotically.
We consider two cases below: firstly, the case where $\psi_{\bm p}(\bm{r})$
is approximated by a plane wave and secondly, the case where $\psi_{\bm p}(\bm{r})$
is an average-atom scattering function that approaches a plane wave asymptotically.

\subsubsection*{Plane-wave final states}

Approximating the final state wave function by a
plane wave $\psi_{\bm{p}}(\bm{r}) = e^{i \bm{p} \cdot \bm{r}}$,
the bound-free structure function in Eq.\ (\ref{snl}) can be rewritten as
\begin{equation}
S_{nl}(k,\omega) =
\frac{o_{nl}}{2l+1}\sum_{m} \int \frac{ p\, d\Omega_p}{(2\pi)^3}
\left| \int\! d^3 r\, e^{i \bm{q}\cdot\bm{r}}\, \psi_{nlm}(\bm{r})
\right|^2
 ,\label{sb}
\end{equation}
where $\epsilon_p = \omega+\epsilon_{nl}$ and
${\bm q}= {\bm k}-{\bm p}$.
Note that ${\bm q}$ is the momentum transferred to the ion. This expression
may be simplified to
\begin{equation}
 S_{nl}(k,\omega) = \frac{o_{nl}}{\pi k} \int_{|p-k|}^{p+k} q\, dq\,
 |K_{nl}(q)|^2, \label{fin}
\end{equation}
\begin{equation}
 K_{nl}(q) = \int_0^\infty\!\! r\, dr\, j_{l}(qr) P_{nl}(r) .
\end{equation}
$S_{nl}(k,\omega)$ in Eq.~(\ref{fin}) depends implicitly on $\omega$
through the relation
\[
p = \sqrt{2(\omega+ \epsilon_{nl}})\ .
\]

%----- figure 5 -----
\begin{figure}[t]
\centering
\includegraphics[scale=0.6]{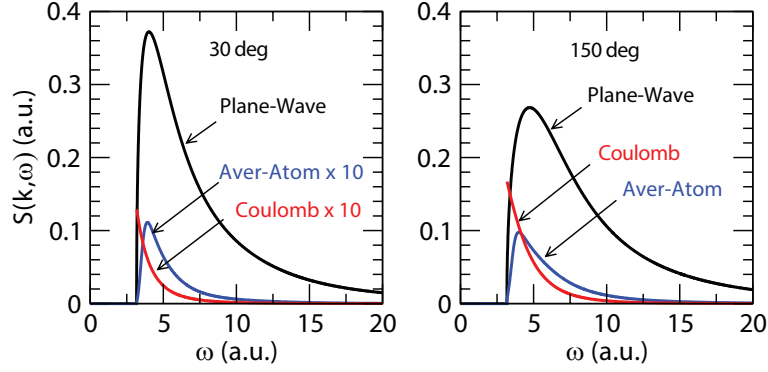}
\caption{The beryllium $K$-shell structure function
$S_{1s}(k,\omega)$ is shown for incident photon energy 2960~eV and
scattering angles 30$^\circ\!$ and 150$^\circ\!$.
The black curves show plane-wave results,
the blue lines show the result obtained using an average-atom
final-state wave function and the red lines show
exact nonrelativistic
Coulomb results. The dramatic suppression of average-atom and Coulomb
structure functions at forward angles (the corresponding curves
are multiplied by 10) is evident in the left panel.\label{fig5}}
\end{figure}
%-----------

\subsubsection*{Average-atom final states}

In the average-atom approach, the final state wave function
consists of a plane wave plus an {\it incoming}
spherical wave. (N.B.\ An outgoing spherical wave is
associated with an incident electron. Time-reversal invariance,
therefore, requires that a converging spherical wave be associated
with an emerging electron.)
The scattering wave function $\psi_{\bm{p}}(\bm{r})$ in Eq.~(\ref{snl})
may be written
\begin{equation}
\psi_{\bm{p}}(\bm{r}) = \frac{4\pi}{p} \sum_{l_1m_1} i^{\, l_1}
e^{-i\delta_{l_1}} \frac{1}{r} P_{\epsilon l_1}(r)\,
Y^\ast_{l_1 m_1}(\hat{p})\, Y_{l_1 m_1}(\hat{r}),
\end{equation}
where the continuum wave function is normalized to a phase-shifted
sine wave asymptotically
\begin{equation}
P_{\epsilon l}(r) \to \sin(pr-l\pi/2 +\delta_l).
\end{equation}
The factor $e^{i\bm{k\cdot r}}$ in Eq.~(\ref{snl}) is expanded as
\begin{equation}
e^{1\bm{k\cdot r}} =  4\pi \sum_{l_2m_2} i^{\, l_2}  j_{l_2}(kr)\,
Y^\ast_{l_2 m_2}(\hat{k})\, Y_{l_2 m_2}(\hat{r}) .
\end{equation}
The bound-free structure function in Eq.~(\ref{snl}) may then be expressed as
\begin{equation}
S_{nl} = \frac{o_{nl}}{2l+1} \int \frac{ p\, d\Omega_p}{(2\pi)^3}
\sum_m \left| J_{nlm} \right|^2,
\end{equation}
where
\begin{equation}
J_{nlm} = \frac{(4\pi)^2}{p} \sum_{l_1 l_2} i^{\, l_2-l_1}\, I_{l_1 l\,l_2}(p,k)
\!\sum_{m_1m_2}\! \left< l_1 m_1 | Y_{lm} | l_2 m_2 \right>
Y_{l_1m_1}(\hat{p})\, Y_{l_2m_2}^\ast(\hat{k}),
\end{equation}
with
\begin{equation}
I_{l_1 l\,l_2}(p,k) = \frac{1}{p}\, e^{i\delta_{l_1}\!\!(p)}
\!\int_0^\infty\! dr P_{\epsilon l_1}(r) P_{nl}(r) j_{l_2}(kr)
\end{equation}
and
\begin{equation}
\left< l_1 m_1 | Y_{lm} | l_2 m_2 \right> =
\int\! d\Omega\, Y^\ast_{l_1m_1}(\hat{r})\, Y_{lm}(\hat{r})\, Y_{l_2m_2}(\hat{r}).
\end{equation}
Squaring $J_{nlm}$, integrating over angles of $\bm{p}$,
and summing over magnetic quantum numbers, one obtains
\begin{equation}
S_{nl} = \frac{2p}{\pi}  o_{nl} \sum_{l_1l_2} A_{l_1 l\,l_2}
\left|I_{l_1 l\,l_2}(p,k)\right|^2 \label{lsum}
\end{equation}
where
\begin{equation}
A_{l_1 l\, l_2} = (2l_1+1)(2l_2+1) \left( \begin{array}{ccc}
l_1 & l & l_2 \\
0 & 0 & 0
\end{array} \right)^{\! 2} .
\end{equation}

In Fig.~\ref{fig5}, several calculations of the structure function
$S_{1s}(k,\omega)$ are compared for a photon of incident energy 2960~eV
scattered at 30$^\circ\!$ ($k=0.411$~a.u.)  and 150$^\circ\!$
($k=1.533$~a.u.) from the $K$ shell of beryllium metal
at $k_{\scriptscriptstyle B} T$ = 20~eV. The results of calculations
carried out using average-atom
final states are smaller than those using plane-wave final states by
a factor of about 40 at $\theta=30^\circ\!$ and 2.5 at $\theta=150^\circ\!$.
This suppression is a characteristic Coulomb field effect.
Indeed, exact nonrelativistic Coulomb-field
calculations of Thomson scattering \cite{EP:70},
with nuclear charge adjusted to align the Coulomb and
average-atom thresholds, illustrated by the curves labeled ``Coulomb''
in Fig.~\ref{fig5}, show a similar suppression.

\begin{figure}[t]
\centering
\includegraphics[scale=0.6]{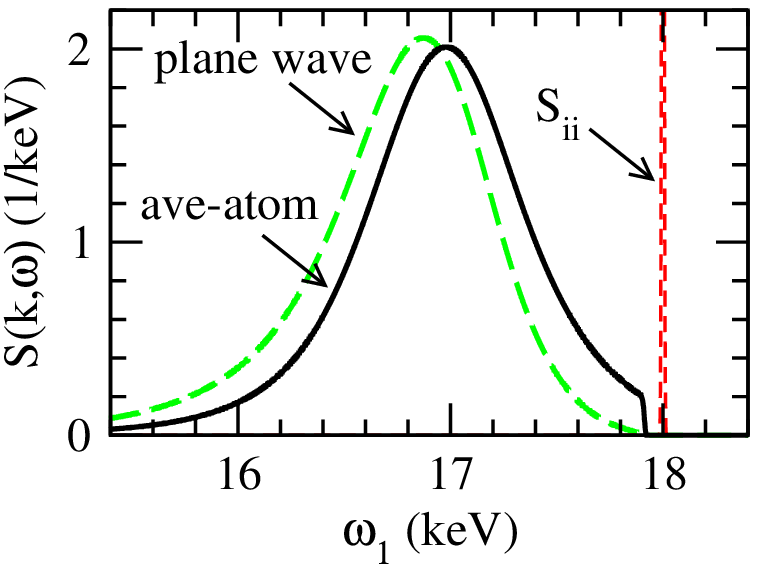}~~
\includegraphics[scale=0.6]{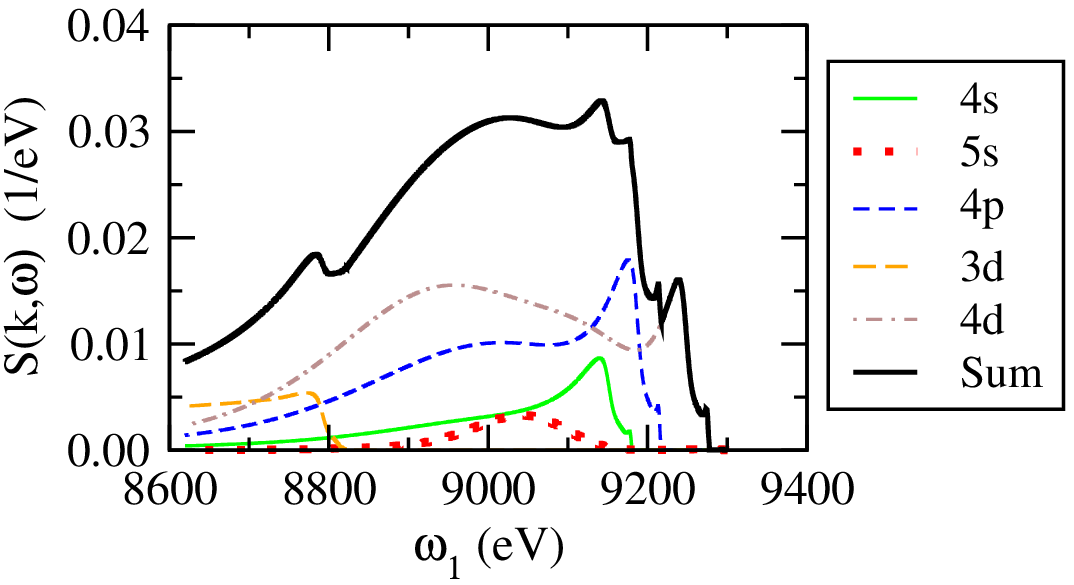}
\caption{Left panel: The dynamic structure function for the $1s$ state of
Be metal at $k_{\scriptscriptstyle B} T = 12$~eV is shown for scattering
of an incident x-ray with energy $\omega_0$ = 18-keV x-ray
at 130$^\circ\!$. The solid black and dashed green lines are results
calculated using average-atom and plane-wave final states, respectively.
The dashed red line is the elastic scattering contribution $S_{ii}(k,\omega)$
with instrumental width 10~eV.
Right panel: The total bound-free dynamic structure function $S_b(k,\omega)$
with average-atom final states is shown for tin ($Z=50$)
at metallic density and $k_{\scriptscriptstyle B} T = 10$~eV is shown,
along with contributions from individual subshells. The incident x-ray energy
is $\omega_0=9.3$~keV and the scattering angle is 130$^\circ\!$.
In the above, $\omega_1 = \omega_0 - \omega$
is the energy of the scattered x-ray. \label{fig6}}
\end{figure}

The situation is quite different at much higher momentum transfer.
This fact is illustrated in the
left panel of Fig.~\ref{fig6}, where we consider scattering of
an 18-keV photon at 130$^\circ\!$ ($k = 8.750$~a.u.)\ from Be metal
at $k_{\scriptscriptstyle B} T$ = 12~eV. The solid black curve shows the
bound-state contribution to the $S(k,\omega)$ evaluated using an average-atom
final state, while the dashed green line shows the contribution
evaluated using a plane-wave final state. In this high-momentum transfer case,
the two calculations differ qualitatively near threshold and the peak energies
differ by a few percent, but the average-atom and plane-wave results are
otherwise quite similar.

In the right panel of Fig.~\ref{fig6}, we consider a more complex example,
scattering of a 9300-eV x-ray at 130$^\circ\!$ ($k=4.521$ a.u.) from tin
metal at $k_{\scriptscriptstyle B} T = 10$~eV.
Contributions from individual subshells are shown together with their sum
$S_b(k,\omega)$.

%----- figure 6 -----
%-----------

\section{Applications \label{apps}}

The scheme developed in the previous sections is applied to several examples
of current or possible future interest. The example of hydrogen at electron density
10$^{24}$~cc$^{-1}$ and temperature 50 eV is considered first. In this example,
the H ion is completely stripped and inelastic scattering is determined completely
by $S_{ee}(k,\omega)$. The plasmon resonances present in the hydrogen spectrum
at forward scattering angles fade rapidly with increasing angle.
The case of beryllium metal at 18~eV, which has been well studied both theoretically
and experimentally, is considered next and good agreement is found between
average-atom predictions and experimental data. For beryllium, the bound-state
contribution is small and beyond the range of available experimental data.
As a third example, calculations of $S(k,\omega)$ for scattering of 9.3-keV and
3.1-keV x-rays from aluminum metal at $k_{\scriptscriptstyle B} T=5$~eV are evaluated
and compared with the average-atom predictions by Sahoo et al.~\cite{SG:08}.
Finally, Thomson scattering of 9.3-keV x-rays from titanium and tin at 30$^\circ\!$
and 130$^\circ\!$ are considered to illustrate cases where bound-state contributions
are the dominant features of the scattered x-ray spectrum. In Table~\ref{tab2},
we list some important average-atom properties for the elements considered in the
following subsections.

%----- table 2 -----
\begin{table}[b]
\centering
\caption{Average-atom parameters for the examples of hydrogen, beryllium,
aluminum, titanium, and tin presented in Sec.~\ref{apps}. \label{tab2}}
\begin{tabular}{l @{\hspace{1pc}}c @{\hspace{1pc}}c @{\hspace{1pc}}c
                  @{\hspace{1pc}}c @{\hspace{1pc}}c}
\hline\hline
& H & Be & Al & Ti & Sn\\
\hline
$k_{\scriptscriptstyle B} T$ (eV)   & 50    &  18    &  5     &  10    & 10     \\
$\rho$   (g/cc)                     & 1.931 &  1.635 &  2.700 &  4.540 &  7.310 \\
$R_{\scriptscriptstyle WS}$~$(a_0)$ & 1.118 &  2.452 &  2.991 &  3.044 &  3.515 \\
$\mu$ (a.u.)                        & 1.091 & -0.531 &  0.241 & -0.044 & -0.071 \\
$N_\text{bound}$                    & 0     &  1.966 & 10.000 & 17.341 & 45.622 \\
$N_\text{cont}$                     & 1     &  2.036 &  3.000 &  4.659 &  4.378 \\
$Z_f$                               & 0.867 &  1.647 &  2.146 &  2.322 &  3.374 \\
$n_e$ ($10^{23}$ cc$^{-1}$)         & 10    &  1.800 &  1.292 &  1.326 &  1.251 \\
\hline\hline
\end{tabular}
\end{table}
%----------

\subsection{Hydrogen}

%----- figure 7 -----
\begin{figure}[t]
\centering
\includegraphics[scale=0.7]{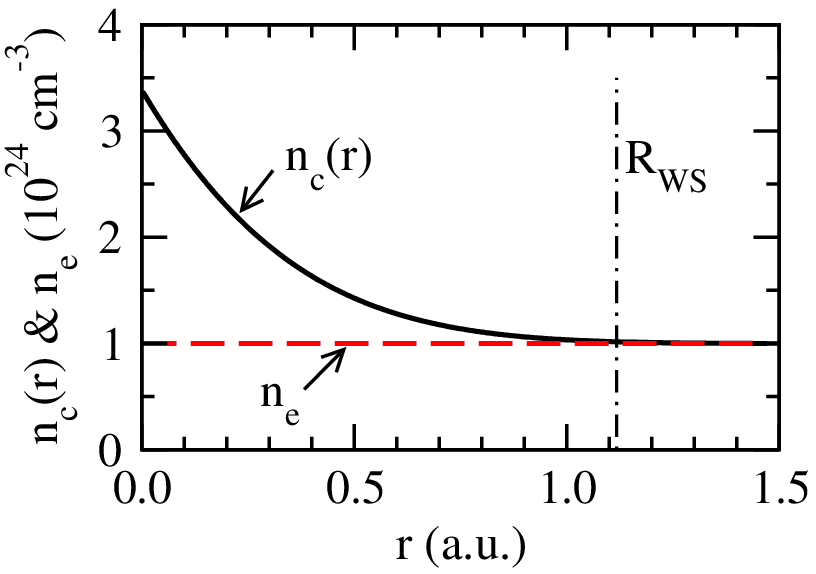}
\includegraphics[scale=0.7]{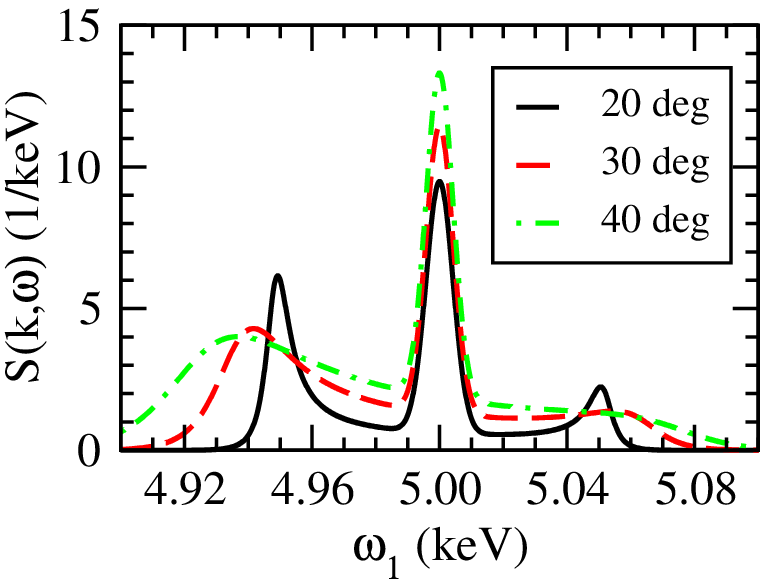}
\caption{Left panel: The continuum density $n_c(r)$ inside the WS sphere
for hydrogen at $k_{\scriptscriptstyle B} T = 50$~eV and density 1.931~g/cc
is seen to converge smoothly to the free-electron density
$n_e = 10^{24}$~cc$^{-1}$ beyond the WS radius $R_{\scriptscriptstyle WS}$.
Right panel: The dynamic structure function $S(k,\omega)$ for H
at $k_{\scriptscriptstyle B}T = 50$~eV and $n_e=10^{24}$~cc$^{-1}$
is plotted for scattering of a 5-keV photon at angles 20$^\circ\!$,
30$^\circ\!$ and 40$^\circ\!$, and $\omega_1 = 5 - \omega$ (keV) is
the energy of the scattered x-ray. \label{fig7} }
\end{figure}
%-----------

In the average-atom model, a density $\rho = 1.931$ g/cc is required at
$k_{\scriptscriptstyle B} T = 50$~eV to achieve free-electron density
$n_e=10^{24}$ cc$^{-1}$. Physical properties of the  plasma under these
conditions are listed in the second column of Table~\ref{tab2}.
The single electron in hydrogen is completely stripped, leaving only the
continuum inside the WS sphere. The continuum density $n_c(r)$ merges
into the uniform free-electron density $n_e$ outside the sphere.
The total number of electrons inside the WS sphere
$N_c =4\pi \int_0^{R_\text{\tiny WS}} r^2 n_c(r) dr =1$,
however, the number of free electrons per ion is $Z_f = 0.867$.
These points are illustrated in the left panel of Fig,~\ref{fig7}.

Since there are no bound electrons, only $S_{ee}(k,\omega)$ contributes
to the inelastic scattering. The dynamic structure functions for scattering
of a 5-keV x-ray at angles 20$^\circ\!$, 30$^\circ\!$ and 40$^\circ\!$
are shown in the right panel of Fig.~\ref{fig7}. The prominent plasmon peaks
at $\theta=20^\circ\!$ wash out at as the scattering angle
(and momentum transfer $k$) is increased.
For this particular case, the screening length
$\lambda_s = 1.071$~a.u.\ differs only slightly
from the WS radius $R_\text{\tiny WS}=1.118$ a.u..
The values of the coherence parameter $\alpha$ for
$\theta=(20^\circ\!,\ 30^\circ\!,\ 40^\circ\!)$ are
$\alpha = (2.005,\ 1.345,\ 1.018)$, respectively.
The resonant features in Fig.~\ref{fig7} are seen to be distinct for $\alpha>1$
but disappear as $\alpha$ approaches 1.
It should be noted that the (unperturbed) plasma frequency for
hydrogen at $n_e = 10^{24}$~cc$^{-1}$ is $\omega_\text{pl} = 37.1$~eV.

\subsection{Beryllium}

\begin{figure}[t]
\centering
\includegraphics[scale=0.7]{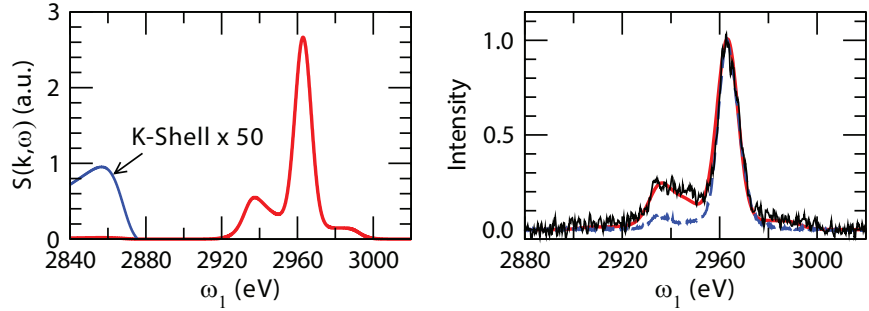}
\caption{Left panel: Structure function $S(k,\omega)$
for scattering of a 2963-eV photon at 40$^\circ$ from
beryllium at $n_e=1.8\times 10^{23}$~cc$^{-1}$ and  electron temperature $18$~eV.
The bound-state contribution is scaled up by a factor of 50.
Right panel: Intensities (arbitrary units) for scattering
of a Cl Ly-$\alpha$ x-ray from beryllium at 40$^\circ\!$ \cite{GD:12};
measured data (black solid line),
source function (blue dashed line),
and the average-atom fit (red solid line).
In the above, $\omega_1$ is the energy of the scattered x-ray.
\label{fig8}}
\end{figure}

In the left panel of Fig.~\ref{fig8}, the dynamic structure function
for scattering of a 2963-eV photon at 40$^\circ\!$ from beryllium
(density = 1.635~g/cc) at $k_{\scriptscriptstyle B} T=18$~eV is shown.
The $L$-shell electrons are completely stripped under these
conditions but the $K$ shell remains 97\% occupied.
The ion temperature, which governs the amplitude of
the elastic peak, is chosen to be $2.1$~eV in this example.
For the case at hand, the coherence parameter is $\alpha = 1.21 >1$,
so one expects and observes plasmon peaks in the scattering intensity profile.
In the average-atom approximation, the threshold energy for bound-state
contributions is the average-atom eigenvalue $\epsilon_{1s}=-86.8$~eV.
(As mentioned earlier, average-atom eigenvalues are inaccurate approximations
to removal energies. The average-atom threshold $|\epsilon_{1s}|$ is 20\%
smaller than the measured $K$-shell threshold 111~eV in beryllium metal \cite{PRD}.)
In the average-atom approximation, bound-state contributions to $S(k,\omega)$
from $K$-shell electrons have a threshold at $\omega_1= 2963-86.8=2876.2$~eV.
The $K$-shell contribution multiplied by 50 is shown in the left panel.
The average-atom parameters used in this calculation are listed in column three
of Table~\ref{tab2}.

To validate the present average-atom model against experimental data,
a Be experiment done at the Omega laser facility that used a Cl Ly-$\alpha$
source to scatter from nearly solid Be at an angle of 40$^\circ$ is used.
An electron temperature of 18~eV, ion temperature of~2.1 eV,
and density of 1.635~g/cc used in the average-atom model
gives an electron density of $1.8 \times 10^{23}$~cc$^{-1}$,
in agreement with the analysis in Ref.~\cite{GD:12}.
The right panel of Fig.~\ref{fig8} shows the experimental source function
from the Cl Ly-$\alpha$ line as a blue dashed line. Because of satellite
structure in the source we approximate the source by three lines:
a Cl Ly-$\alpha$ line at 2963~eV with amplitude 1 and two satellites
at 2934 and 2946~eV with relative amplitudes of 0.075 and 0.037, respectively.
Doing the Thomson scattering calculation using the three weighted lines,
we calculate the scattering amplitude for Thomson scattering
(red solid line)
and compare against the experimental data (black solid line) here.
We observe excellent agreement within the experimental noise.
Contributions from the bound $1s$ electrons, which have a threshold at 2876~eV,
are beyond the range of the data shown in the right panel.

%----- figure 8 -----
%-----------

\subsection{Aluminum}

\begin{figure}[t]
\centering
\includegraphics[scale=0.8]{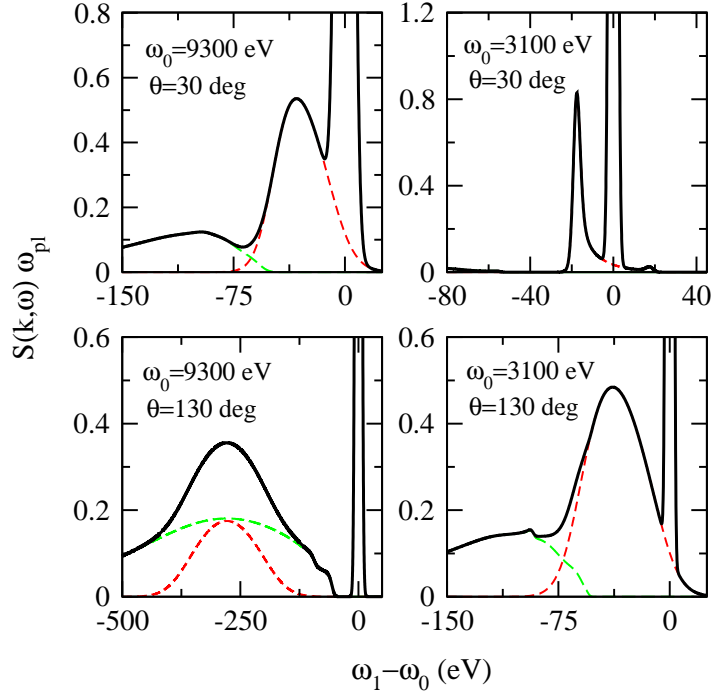}
\caption{$S(k,\omega)\, \omega_\text{pl}$ is shown for scattering of
9300-eV and 3100-eV x-rays at 30$^\circ\!$ and 130$^\circ\!$ from Al metal
at $k_{\scriptscriptstyle B} T$ = 5~eV. The short-dashed red curves
are contributions $S_{ee}(k,\omega)$ from
free electrons, the long-dashed green curves give contributions $S_b(k,\omega)$
from bound $2s$ and $2p$ electrons and the solid black curves show the total
dynamic structure function. Instrumental widths of 0.1\%$\omega_0$ were used.
For Al metal at $k_{\scriptscriptstyle B} T$ = 5~eV, the plasma frequency
is $\omega_\text{pl} = 13.35$~eV. In the above figure, $\omega_1$ and
$\omega_0$ are energies of the scattered
and incident x-rays, respectively,
and $\omega = \omega_0 - \omega_1$. \label{fig9}}
\end{figure}

As mentioned in the introduction, the present average-atom calculations disagree
in various ways with those in Ref.~\cite{SG:08}, where Thomson scattering
from aluminum metal at several temperatures was considered.
To clarify the differences, we compare results of the present calculations
for the case of aluminum at $k_{\scriptscriptstyle B} T = 5$~eV
with the corresponding results presented in Ref.~\cite{SG:08}.
The dimensionless product $S(k,\omega)\, \omega_\text{pl}$ is shown for
scattering of 9300-eV and 3100-eV x-rays
in the left and right panels of Fig.~\ref{fig9}, respectively. The upper and lower
panels of Fig.~\ref{fig9} show results for scattering angles of 30$^\circ\!$ and
130$^\circ\!$, respectively. The plots in Fig.~\ref{fig9} can be compared directly
with those in Fig.~6 and Fig.~7 of \cite{SG:08}, where similar plots for 9300-eV
and 3100-eV x-ray energies can be found.
The size and shape of the free-electron contributions shown in Fig.~\ref{fig9}
are in agreement with those shown in the corresponding figures in \cite{SG:08},
however, the $L$-shell contributions to $S(k,\omega)$ in \cite{SG:08} are larger
by a factor of approximately 5 than those shown in Fig.~\ref{fig9}.
Moreover, contributions from the $M$ shell, which dominate the spectrum
just below the elastic peak in \cite{SG:08} are completely absent
in Fig.~\ref{fig9}. Owing to the differences in bound-free
contributions $S_b(k,\omega)$, the present prediction for the aluminum
dynamic structure function $S(k,\omega)$ differs in both size and shape
from that given in \cite{SG:08}.

Differences in boundary conditions between the two average-atom models
account for the presence or
absence of bound $M$-shell electrons as discussed in Sec.~\ref{avat}.
Furthermore, difference in the relative size of the $L$-shell contributions
at forward angles may owe in part to the use of plane-wave final states
in \cite{SG:08}.
However, the substantial difference in size of the $L$-shell contributions
at backward angles remains a mystery.
The average-atom parameters for aluminum used in the present calculation
are given in the fourth column of Table~\ref{tab2}.

%----- figure 9 -----
%-----------

\subsection{Titanium \& Tin}

%----- figure 10 -----
\begin{figure}[t]
\centering
\includegraphics[scale=0.8]{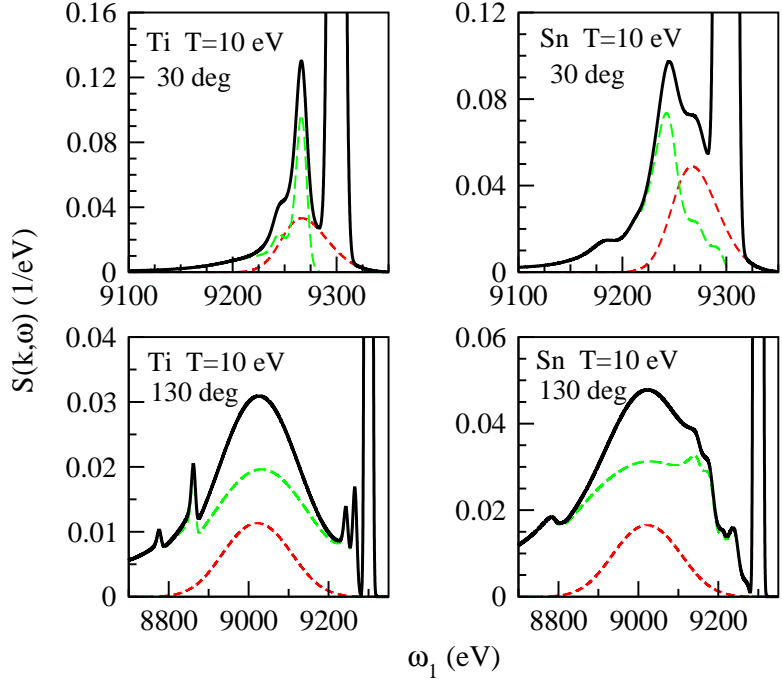}
\caption{Left panels: $S(k,\omega)$ for scattering of a 9.3-keV x-ray at
$\theta = 30^\circ\!$ and 130$^\circ\!$ from Ti metal
at $k_{\scriptscriptstyle B}=10$~eV is shown in the solid black
line. The free-electron contribution $S_{ee}(k,\omega)$
is shown in the short-dashed red line while the contribution
from bound states $S_b(k,\omega)$ is shown in the
long-dashed green line.
Right panel: Same as left panel, except scattering from Sn metal at
$k_{\scriptscriptstyle B}=10$~eV. Contributions from individual subshells
of Sn are shown in the right panel of Fig.~\ref{fig10}.
In the figure above, $\omega_1 = 9300 -\omega$ (eV)
is the energy of the scattered x-ray.
\label{fig10}}
\end{figure}
%-----------

%----- table 3 -----
\begin{table}[b]
\centering
\caption{Comparison of the theoretical average-atom thresholds
($-\epsilon_{nl}$) and experimental thresholds from the
NIST database \cite{PRD} for contributions to $S_b(k,\omega)$
from individual subshells in titanium and tin.  \label{tab3} }
\begin{tabular}{c @{\hspace{1pc}}r @{\hspace{1pc}}r @{\hspace{1pc}}r
                  @{}c@{\hspace{1.5pc}}
                c @{\hspace{1pc}}r @{\hspace{1pc}}r @{\hspace{1pc}}r}
\hline\hline
\multicolumn{4}{c}{Titanium} && \multicolumn{4}{c}{Tin} \\
\cline{1-4} \cline{6-9}
$nl$ & theory & expt. & $\Delta$\% &&
$nl$ & theory & expt. & $\Delta$\% \\
\hline
$2s$ & 512.7 & 563.7 &  9   &&   $3d$ & 477.8 & 488.2 &  2  \\
$2p$ & 426.8 & 457.5 &  7   &&   $4s$ & 118.9 & 136.5 & 13  \\
$3s$ &  45.4 &  60.3 & 25   &&   $4p$ &  83.1 &  86.6 &  4  \\
$3p$ &  22.8 &  34.6 & 34   &&   $4d$ &  22.4 &  23.9 &  6  \\
     &       &       &      &&   $5s$ &   2.1 &  --~~ & --  \\
\hline\hline
\end{tabular}
\end{table}
%----------

The average-atom model predicts that a titanium metal at 10~eV is in an Ar-like
configuration with completely filled $K$ and $L$ shells together with 1.97 $3s$
electrons  5.36  $3p$ electrons in the $M$ shell.
This is an interesting case where contributions to $S_b(k,\omega)$
from the $L$ and $M$ shells are substantial.
In the left panels of Fig.~\ref{fig10}, the dynamic structure function
$S(k,\omega)$ is shown in the solid black line for an incident 9.3-keV x-ray
scattered at 30$^\circ\!$ and 130$^\circ\!$.
Contributions from $S_b(k,\omega)$ shown in the long-dashed green lines
dominate the free-electron contributions $S_{ee}(k,\omega)$ shown in the
short-dashed red lines. As seen in the lower-left panel,
sharp features associated with excitation of the $M$ subshells ($3p$ and $3s$)
show up just below the elastic peak at $\omega_1 = \omega_0$,
whereas features associated with excitation of the $L$ subshells ($2p$ and $2s$)
show up 400 to 500 eV below the elastic peak.
The bound-state thresholds
for titanium are compared with measured thresholds from the National Institute
for Standards \& Technology (NIST) database \cite{PRD}
in Table~\ref{tab3}.

Metallic tin at 10 eV has a Pd configuration with 45.6 bound electrons.
The resulting dynamic structure function $S(k,\omega)$ is shown in the right panels
of Fig.~\ref{fig10} for scattering angles of 30$^\circ\!$ and 130$^\circ\!$.
The situation is similar to that for titanium;
bound-free contributions dominate those from free-electrons.
The irregularities in the bound-state contributions are associated
with the onset of contributions from individual subshells.
For scattering at 130$^\circ\!$, contributions from individual subshells
are shown in the right panel of Fig.~\ref{fig6}.
Average-atom thresholds for tin are compared with values from the NIST
database \cite{PRD} in Table~\ref{tab3}.

\section{Summary}

A scheme for analysis of Thomson scattering from plasmas based
on the average-atom model is presented.
Given the plasma composition $(Z,A)$, density $\rho$ and temperature $T$,
the model gives, in addition to the equation of state of the plasma,
all parameters needed for a complete description of the Thomson scattering
process. In particular, the average-atom code predicts wave functions
for bound and continuum electrons, densities of bound, screening,
and free electrons, and the chemical potential.
Predictions of the present average-atom model disagree with those in
Ref.~\cite{SG:08} where a similar model with different boundary conditions
was used. In the average-atom model used in Ref.~\cite{SG:08},
$3d$ ($M$ shell) electrons are bound in metallic Al
for temperatures between 2 and 10~eV, leading to substantial bound-state
contributions to the dynamic structure function. In the present model,
by contrast, the $M$ shell of metallic aluminum is vacant in the
temperature range $k_{\scriptscriptstyle B}T\leq 10$~eV and the
corresponding bound-state features are absent.

Elastic scattering from bound and screening electrons is treated here
following the model proposed by Gergori et al.~\cite{GG:03} which makes
use of formulas for the static ion-ion structure function $S_{ii}(k)$
given by Arkhipov and Davletov~\cite{AD:98}.
Modifications suggested in Ref.~\cite{GGL:06} to account for different
electron and ion temperatures are also included. Specifically,
in the applications considered here, the amplitude of the elastic peak
is adjusted artificially by choosing an ion temperature that is different
from the electron temperature, even in cases where equilibrium is expected.
Such an adjustment was
used to fit the experimental data for beryllium shown in Fig.~\ref{fig8}.
The dynamic structure function for scattering from free electrons depends
sensitively on the free-electron dielectric function $\epsilon(k,\omega)$.
Again, we follow the model proposed in Ref.~\cite{GG:03} and evaluate
the dielectric function in the random-phase approximation. The RPA dielectric
function includes features such as plasmon resonant peaks that show up in
experimental intensity profiles and can be used in connection
with the principle of detailed balance to determine electron temperatures.
Bound-state features are easily included in the present scheme, inasmuch
as the average-atom model provides bound-state and continuum wave functions.
Ionic Coulomb-field effects are features of calculations
carried out using average-atom wave functions rather than plane waves
to describe the final state electrons.
In conclusion, the average-atom model provides a simple and consistent point
of departure for the theoretical analysis of Thomson scattering from plasmas.

\section*{Acknowledgements}
We owe debts of gratitude to S. H. Glenzer, C. Fortmann and T. D\"{o}ppner
for informative discussions and for providing comparison experimental data
for x-ray scattering from beryllium.
The work of J.N.\ and K.T.C.\ was performed under the auspices of the U.S.\
Department of Energy by Lawrence Livermore National Laboratory
under Contract DE-AC52-07NA27344.

%\bibliography{xbib}

\end{document}